\documentclass[acmsmall,screen]{acmart}

\newcommand{\reviewAddr}[1]{}

\newcommand{\rqa}{$RQ_{1}$}
\newcommand{\rqb}{$RQ_{2}$}
\newcommand{\rqc}{$RQ_{3}$}

\newcommand{\rqaa}{How accurately can we identify when vulnerabilities were introduced?}
\newcommand{\rqbb}{What metrics are effective for vulnerability prediction?}
\newcommand{\rqcc}{\reviewAddr{G-1} Which allocation strategy is most effective for fuzzing?}

\newcommand{\rqA}{\rqa: \rqaa}
\newcommand{\rqB}{\rqb: \rqbb}
\newcommand{\rqC}{\rqc: \rqcc}
\newcommand{\app}{Risk-Aware Intensive Supplemental Execution Framework\xspace}
\newcommand{\APP}{\textsl{RAISE}\xspace}

\newcommand{\ccs}{code changes}

\newcommand{\currentTargetProject}{1,320\xspace}   

\newcommand{\ccppvulnerabilities}{3,512\xspace}
\newcommand{\ccppprojects}{268\xspace}
\newcommand{\ccpppercent}{91.9} 
\newcommand{\clonedvulnerabilities}{3,439\xspace}
\newcommand{\clonedprojects}{255\xspace}
\newcommand{\finalprojects}{15\xspace}
\newcommand{\finalpositivedays}{347\xspace}
\newcommand{\finalnegativedays}{\num{14863}\xspace}
\newcommand{\finalpositivedaysnum}{347}
\newcommand{\finalnegativedaysnum}{14863}
\newcommand{\finaltotaldaysnum}{\fpeval{\finalpositivedaysnum + \finalnegativedaysnum}} 

\newcommand{\finaltotaldays}{\num{\finaltotaldaysnum}} 
\newcommand{\finaldaysratio}{\fpeval{round(100 * \finalpositivedaysnum / \finaltotaldaysnum,2)}\%\xspace}

\newcommand{\finalcoveragereport}{\num{30312}\xspace} 
\newcommand{\finalsrcmap}{\num{29879}\xspace} 

\newcommand{\scrapestartdate}{August 22, 2018\xspace} 
\newcommand{\scrapeenddate}{June 1, 2025\xspace} 

\newcommand{\allvulnerabilityreport}{3,822\xspace}
\newcommand{\numberofvulnerabilities}{968\xspace}

\newcommand{\typeofcoveragefeature}{7\xspace}

\newcommand{\tworesearchfeature}{77\xspace}
\newcommand{\allfeature}{88\xspace}

\newcommand{\bestmodel}{Random forest\xspace}
\newcommand{\bestperfomancefeature}{VCCFinder + Coverage\xspace}
\newcommand{\bestperfomanceexp}{fs4\xspace}

\newcommand{\expZeroRandomMccValuenum}{0.00}
\newcommand{\expZeroRandomMccValue}{\expZeroRandomMccValuenum \xspace}
\newcommand{\expZeroRandomFoneValuenum}{0.02}
\newcommand{\expZeroRandomFoneValue}{\expZeroRandomFoneValuenum \xspace}
\newcommand{\expZeroRandomPrAucValuenum}{0.03}
\newcommand{\expZeroRandomPrAucValue}{\expZeroRandomPrAucValuenum \xspace}
\newcommand{\expZeroRandomRecallValuenum}{0.02}
\newcommand{\expZeroRandomRecallValue}{\expZeroRandomRecallValuenum \xspace}
\newcommand{\expZeroRandomPrecisionValuenum}{0.03}
\newcommand{\expZeroRandomPrecisionValue}{\expZeroRandomPrecisionValuenum \xspace}
\newcommand{\expZeroRandomRocAucValuenum}{0.50}
\newcommand{\expZeroRandomRocAucValue}{\expZeroRandomRocAucValuenum \xspace}

\newcommand{\expOneRfMccValuenum}{0.04}

\newcommand{\expOneRfFoneValuenum}{0.07}

\newcommand{\expOneRfPrAucValuenum}{0.05}
\newcommand{\expOneRfPrAucValue}{\expOneRfPrAucValuenum \xspace}
\newcommand{\expOneRfRecallValuenum}{0.51}

\newcommand{\expOneRfPrecisionValuenum}{0.04}

\newcommand{\expOneRfRocAucValuenum}{0.60}
\newcommand{\expOneRfRocAucValue}{\expOneRfRocAucValuenum \xspace}

\newcommand{\expOneXgbMccValuenum}{0.04}

\newcommand{\expOneXgbFoneValuenum}{0.07}

\newcommand{\expOneXgbPrAucValuenum}{0.05}

\newcommand{\expOneXgbRecallValuenum}{0.51}

\newcommand{\expOneXgbPrecisionValuenum}{0.04}

\newcommand{\expOneXgbRocAucValuenum}{0.59}

\newcommand{\expTwoRfMccValuenum}{0.11}

\newcommand{\expTwoRfFoneValuenum}{0.11}

\newcommand{\expTwoRfPrAucValuenum}{0.10}
\newcommand{\expTwoRfPrAucValue}{\expTwoRfPrAucValuenum \xspace}
\newcommand{\expTwoRfRecallValuenum}{0.62}

\newcommand{\expTwoRfPrecisionValuenum}{0.06}

\newcommand{\expTwoRfRocAucValuenum}{0.71}
\newcommand{\expTwoRfRocAucValue}{\expTwoRfRocAucValuenum \xspace}

\newcommand{\expTwoXgbMccValuenum}{0.10}

\newcommand{\expTwoXgbFoneValuenum}{0.10}

\newcommand{\expTwoXgbPrAucValuenum}{0.10}

\newcommand{\expTwoXgbRecallValuenum}{0.63}

\newcommand{\expTwoXgbPrecisionValuenum}{0.06}

\newcommand{\expTwoXgbRocAucValuenum}{0.70}

\newcommand{\expThreeRfMccValuenum}{0.08}
\newcommand{\expThreeRfMccValue}{\expThreeRfMccValuenum \xspace}
\newcommand{\expThreeRfFoneValuenum}{0.09}

\newcommand{\expThreeRfPrAucValuenum}{0.06}
\newcommand{\expThreeRfPrAucValue}{\expThreeRfPrAucValuenum \xspace}
\newcommand{\expThreeRfRecallValuenum}{0.66}

\newcommand{\expThreeRfPrecisionValuenum}{0.05}

\newcommand{\expThreeRfRocAucValuenum}{0.66}

\newcommand{\expThreeXgbMccValuenum}{0.07}

\newcommand{\expThreeXgbFoneValuenum}{0.08}

\newcommand{\expThreeXgbPrAucValuenum}{0.06}

\newcommand{\expThreeXgbRecallValuenum}{0.66}

\newcommand{\expThreeXgbPrecisionValuenum}{0.05}

\newcommand{\expThreeXgbRocAucValuenum}{0.65}

\newcommand{\expFourRfMccValuenum}{0.12}
\newcommand{\expFourRfMccValue}{\expFourRfMccValuenum \xspace}
\newcommand{\expFourRfFoneValuenum}{0.11}
\newcommand{\expFourRfFoneValue}{\expFourRfFoneValuenum \xspace}
\newcommand{\expFourRfPrAucValuenum}{0.10}
\newcommand{\expFourRfPrAucValue}{\expFourRfPrAucValuenum \xspace}
\newcommand{\expFourRfRecallValuenum}{0.66}
\newcommand{\expFourRfRecallValue}{\expFourRfRecallValuenum \xspace}
\newcommand{\expFourRfPrecisionValuenum}{0.06}

\newcommand{\expFourRfRocAucValuenum}{0.74}
\newcommand{\expFourRfRocAucValue}{\expFourRfRocAucValuenum \xspace}

\newcommand{\expFourXgbMccValuenum}{0.11}

\newcommand{\expFourXgbFoneValuenum}{0.11}

\newcommand{\expFourXgbPrAucValuenum}{0.11}

\newcommand{\expFourXgbRecallValuenum}{0.68}

\newcommand{\expFourXgbPrecisionValuenum}{0.06}

\newcommand{\expFourXgbRocAucValuenum}{0.72}

\newcommand{\expFiveRfMccValuenum}{0.08}

\newcommand{\expFiveRfFoneValuenum}{0.09}

\newcommand{\expFiveRfPrAucValuenum}{0.09}

\newcommand{\expFiveRfRecallValuenum}{0.64}

\newcommand{\expFiveRfPrecisionValuenum}{0.05}

\newcommand{\expFiveRfRocAucValuenum}{0.69}

\newcommand{\expFiveXgbMccValuenum}{0.09}

\newcommand{\expFiveXgbFoneValuenum}{0.09}

\newcommand{\expFiveXgbPrAucValuenum}{0.09}

\newcommand{\expFiveXgbRecallValuenum}{0.66}

\newcommand{\expFiveXgbPrecisionValuenum}{0.05}

\newcommand{\expFiveXgbRocAucValuenum}{0.68}

\newcommand{\SoneDetectionVulnerabilityRateNum}{2.32} 
\newcommand{\SoneDetectionVulnerabilityRate}{\SoneDetectionVulnerabilityRateNum}
\newcommand{\SoneAdditionalBuildsTotalNum}{35925} 
\newcommand{\SoneAdditionalBuildsTotal}{\num{\SoneAdditionalBuildsTotalNum}\xspace}

\newcommand{\SoneBuildsPerDetectionNum}{171.89} 
\newcommand{\SoneBuildsPerDetection}{\SoneBuildsPerDetectionNum\xspace}
\newcommand{\SoneFoundVulnsNum}{209} 
\newcommand{\SoneFoundVulns}{\SoneFoundVulnsNum\xspace}

\newcommand{\StwoDetectionVulnerabilityRateNum}{2.78} 
\newcommand{\StwoDetectionVulnerabilityRate}{\StwoDetectionVulnerabilityRateNum}
\newcommand{\StwoAdditionalBuildsTotalNum}{188985} 
\newcommand{\StwoAdditionalBuildsTotal}{\num{\StwoAdditionalBuildsTotalNum}\xspace}

\newcommand{\StwoBuildsPerDetectionNum}{679.80} 
\newcommand{\StwoBuildsPerDetection}{\StwoBuildsPerDetectionNum\xspace}
\newcommand{\StwoFoundVulnsNum}{278} 
\newcommand{\StwoFoundVulns}{\StwoFoundVulnsNum\xspace}

\newcommand{\SthreeDetectionVulnerabilityRateNum}{2.32} 
\newcommand{\SthreeDetectionVulnerabilityRate}{\SthreeDetectionVulnerabilityRateNum}
\newcommand{\SthreeAdditionalBuildsTotalNum}{38334} 
\newcommand{\SthreeAdditionalBuildsTotal}{\num{\SthreeAdditionalBuildsTotalNum}\xspace}

\newcommand{\SthreeBuildsPerDetectionNum}{179.97} 
\newcommand{\SthreeBuildsPerDetection}{\SthreeBuildsPerDetectionNum\xspace}
\newcommand{\SthreeFoundVulnsNum}{213} 
\newcommand{\SthreeFoundVulns}{\SthreeFoundVulnsNum\xspace}

\newcommand{\SfourDetectionVulnerabilityRateNum}{1.72} 
\newcommand{\SfourDetectionVulnerabilityRate}{\SfourDetectionVulnerabilityRateNum}
\newcommand{\SfourAdditionalBuildsTotalNum}{11084} 
\newcommand{\SfourAdditionalBuildsTotal}{\num{\SfourAdditionalBuildsTotalNum}\xspace}

\newcommand{\SfourBuildsPerDetectionNum}{91.60} 
\newcommand{\SfourBuildsPerDetection}{\SfourBuildsPerDetectionNum\xspace}
\newcommand{\SfourFoundVulnsNum}{121} 
\newcommand{\SfourFoundVulns}{\SfourFoundVulnsNum\xspace}

\newcommand{\SfiveDetectionVulnerabilityRateNum}{2.34} 
\newcommand{\SfiveDetectionVulnerabilityRate}{\SfiveDetectionVulnerabilityRateNum}
\newcommand{\SfiveAdditionalBuildsTotalNum}{18861} 
\newcommand{\SfiveAdditionalBuildsTotal}{\num{\SfiveAdditionalBuildsTotalNum}\xspace}

\newcommand{\SfiveBuildsPerDetectionNum}{176.27} 
\newcommand{\SfiveBuildsPerDetection}{\SfiveBuildsPerDetectionNum\xspace}
\newcommand{\SfiveFoundVulnsNum}{107} 
\newcommand{\SfiveFoundVulns}{\SfiveFoundVulnsNum\xspace}

\newcommand{\deltaFourThreeRocAucValuenum}{0.70} 
\newcommand{\deltaFourThreeRocAucValue}{\deltaFourThreeRocAucValuenum\xspace}
\newcommand{\deltaFourTwoRocAucValuenum}{0.27} 
\newcommand{\deltaFourTwoRocAucValue}{\deltaFourTwoRocAucValuenum\xspace}
\newcommand{\deltaSFourSFiveValuenum}{0.43} 
\newcommand{\deltaSFourSFiveValue}{\deltaSFourSFiveValuenum\xspace}
\newcommand{\pvalSFourSFiveValuenum}{0.004} 
\newcommand{\pvalSFourSFiveValue}{\pvalSFourSFiveValuenum\xspace}

\usepackage{fancyvrb}
\usepackage{xurl}
\usepackage{booktabs}
\usepackage{url}
\usepackage{amsmath,amsfonts}
\usepackage{algorithmic}
\usepackage{graphicx}
\usepackage{textcomp}
\usepackage{hyperref}
\usepackage{makecell}
\usepackage{xspace}
\usepackage{subcaption}
\usepackage{tabularx}
\usepackage{amsmath}
\usepackage{balance}
\usepackage{framed}
\usepackage{chngpage}
\usepackage[many]{tcolorbox}

\usepackage{graphicx}

\usepackage{pifont}

\usepackage{enumitem}

\usepackage{xfp} 

\usepackage{multirow} 
\usepackage{listings}

\usepackage[group-separator={,}, group-minimum-digits=4]{siunitx}

\definecolor{darkgreen}{rgb}{0, 0.5, 0} 
\definecolor{whitesmoke}{rgb}{0.99, 0.99, 0.99} 

\def\Underline{\setbox0\hbox\bgroup\let\\\endUnderline}
\def\endUnderline{\vphantom{y}\egroup\smash{\underline{\box0}}\\}
\def\|{\verb|}

\newcommand{\ie}{\textit{i.e.,}\xspace}
\newcommand{\eg}{\textit{e.g.,}\xspace}

\newcommand{\etal}{\xspace\textit{et al.}\xspace}

\newcounter{findings_no}

\usepackage{listings}

\definecolor{backcolour}{rgb}{0.95,0.95,0.92}
\lstdefinelanguage{diff}{
  morecomment=**[f][\color{red}]{-},         
  morecomment=**[f][\color{darkgreen}]{+},       
  moredelim=**[is][\bfseries]{@@}{@@},
}
\definecolor{backcolour}{rgb}{0.95,0.95,0.92}
\lstdefinelanguage{commit}{ 
  breakindent = 0pt,
  numbers=none,
  backgroundcolor=\color{white},
  frame=single,
  xleftmargin=3.5em,
  numbersep=0em,
  xrightmargin=1.5em,
}

\usepackage[many]{tcolorbox}  
\tcbuselibrary{listings,breakable}
\definecolor{main}{HTML}{D0D3D4}    
\definecolor{sub}{HTML}{D0D3D4}     
\newtcolorbox{dbox}{
    left=2pt,right=2pt,top=2pt,bottom=2pt,
    enhanced, 
    boxrule = 0pt,
    enlarge top by=5pt,
    enlarge bottom by=3pt,
  }
\tcbset{
    sharp corners,
    before skip = 0.1cm,    
    after skip = 0.01cm      
}

\def\summarybox#1#2{
\smallskip
\begin{tcolorbox}[
  enhanced,
  title=#1,
  colframe=darkgray,
  boxsep=2pt,        
  left=4pt,          
  right=4pt,         
  top=2pt,           
  bottom=2pt,        
]
    #2
\end{tcolorbox}
}
\DeclareUnicodeCharacter{2212}{-}

\setcopyright{acmlicensed}
\copyrightyear{2026}
\acmYear{2026}
\acmISBN{978-1-4503-XXXX-X/18/06}

\begin{document}

\title{Raise or Check? Predicting the Necessity of Additional Fuzzing for Continuous Fuzzing}

\author{Riku Kato}
\affiliation{%
  \institution{Nara Institute of Science and Technology}
  \city{Ikoma}
  \country{Japan}
}
\email{kato.riku.ks5@naist.ac.jp}

\author{Tatsuya Shirai}
\affiliation{%
  \institution{Nara Institute of Science and Technology}
  \city{Ikoma}
  \country{Japan}
}
\email{shirai.tatsuya.sp1@naist.ac.jp}

\author{Olivier Nourry}
\affiliation{%
  \institution{The University of Osaka}
  \city{Suita}
  \country{Japan}
}
\email{nourry@ist.osaka-u.ac.jp}

\author{Yutaro Kashiwa}
\affiliation{%
  \institution{Nara Institute of Science and Technology}
  \city{Ikoma}
  \country{Japan}
}
\email{yutaro.kashiwa@is.naist.jp}

\author{Kenji Fujiwara}
\affiliation{%
  \institution{Nara Women's University}
  \city{Nara}
  \country{Japan}
}
\email{kenjif@ics.nara-wu.ac.jp}

\author{Yasutaka Kamei}
\affiliation{%
  \institution{Kyushu University}
  \city{Fukuoka}
  \country{Japan}
}
\email{kamei@ait.kyushu-u.ac.jp}

\author{Hajimu Iida}
\affiliation{%
  \institution{Nara Institute of Science and Technology}
  \city{Ikoma}
  \country{Japan}
}
\email{iida@naist.ac.jp}

\renewcommand{\shortauthors}{Kato et al.}




\begin{abstract}
Continuous fuzzing is used for vulnerability detection in Open Source Software (OSS). Large-scale services such as Google’s OSS-Fuzz fuzz over 1,300 projects and have revealed more than 10,000 vulnerabilities.
Despite the widespread adoption of continuous fuzzing, a considerable time lag often exists between when a vulnerability is introduced and when it is detected (\ie the window of exposure). 
Given the computational cost of additional fuzzing, we aim to allocate fuzzing resources efficiently using vulnerability-risk estimates.
We propose \APP (\app), a Just-In-Time (JIT) framework that prioritizes additional fuzzing for high-risk project-days.
\APP performs binary classification at the project-day granularity to estimate whether a vulnerability is introduced on a given day. We construct a dataset by combining vulnerability-introducing commits from the OSV database with daily coverage and change metrics from OSS-Fuzz, covering 15 C/C++ projects and 15,210 project-days, of which 347 project-days (2.28\%) are labeled as vulnerability-introducing. 

We evaluate \APP in a cross-project validation setting and use a simulation framework to compare five strategies for allocating additional fuzzing sessions.
Our best model, a Random Forest classifier combining VCCFinder-based static features with dynamic fuzzing features, achieves a ROC-AUC of 0.74, substantially outperforming a random baseline. Coverage metrics receive the highest importance scores, with PatchCoverage ranking first. In a simulation over 968 vulnerabilities, the Simple-Regression strategy was significantly more cost-efficient than the prediction-agnostic heuristics, spending \SfourBuildsPerDetection additional fuzzing sessions per detected vulnerability, roughly half the cost of the Median heuristic (\SoneBuildsPerDetection sessions).
\APP demonstrates that JIT vulnerability prediction augmented with coverage-aware fuzzing features can effectively guide risk-aware allocation of continuous fuzzing resources, enabling more efficient use of computation 
in large-scale OSS fuzzing environments.

\end{abstract}


\maketitle

\section{Introduction}
\label{sec:Introduction} 
Numerous large-scale security incidents have recently exploited vulnerabilities in Open Source Software (OSS), including well-known cases such as Heartbleed \cite{Heartbleed} and RegreSSHion \cite{regReSSHion}. In modern software development, OSS components are extensively reused across a wide range of software projects, including other OSS projects,  thereby amplifying the potential impact of software vulnerabilities. Early identification of vulnerabilities prior to their exploitation is therefore an essential aspect of software security maintenance. 

Among the various techniques proposed for vulnerability detection, fuzz testing (or fuzzing) has emerged as a particularly promising approach in recent literature~\cite{DBLP:journals/cybersec/LiZZ18,DBLP:journals/tse/ManesHHCESW21,DBLP:journals/compsec/BeamanRMH22}.
Fuzzing is a dynamic analysis technique that systematically generates and sends malformed or unexpected data to software systems with the goal of provoking anomalous behaviors such as crashes or exceptions that may indicate underlying security flaws.
Cui\etal\cite{DBLP:journals/compsec/CuiCHLDL22} reported that fuzzing can discover more previously unknown vulnerabilities than other vulnerability discovery techniques, such as static analysis and concolic testing. 
They also surveyed 57 papers focusing on fuzzing techniques, collected 1,600 confirmed vulnerabilities, and found that 635 (39.7\%) of them were assigned Common Vulnerabilities and Exposures (CVE) IDs. Their field survey results also showed that only 13.2\% of vulnerabilities detected using static-approaches were later assigned a CVE ID, suggesting that vulnerabilities found via fuzzing are more likely to be assigned a CVE ID.

Fuzzing also has its drawbacks. Specifically, it is known to require significant execution time, as it involves generating and executing a large number of inputs against the target software system \cite{DBLP:conf/sigsoft/BohmeF20, DBLP:journals/tosem/NourryKLBLK24}. 
To mitigate the drawbacks of traditional fuzzing, continuous fuzzing integrates directly into the CI pipeline, running in short, frequent bursts rather than lengthy sessions. This approach minimizes the delay between code commits and fuzzing results, enabling developers to catch vulnerabilities closer to when they are introduced.  
Since 2016, Google has been conducting large-scale fuzzing through OSS-Fuzz~\cite{203944}, a service that performs continuous fuzzing on OSS to improve security. As of 2023, OSS-Fuzz has targeted over 1,000 projects and discovered more than 10,000 vulnerabilities and 36,000 bugs~\cite{OSSFuzz_Document}.

Continuous fuzzing, however, does not accelerate the fuzzer’s inherent search process. Therefore, even with continuous fuzzing, significant time may pass between when a vulnerability is introduced and when it is discovered, leaving the code exposed to long-term exploitation. 
In fact, prior work has reported detection latencies under continuous fuzzing. 
For example, Keller\etal\cite{DBLP:conf/msr/KellerMM23} found that in OSS-Fuzz, the median time from bug introduction to fix was 324 days, despite fixes typically being applied within 2 days of detection. 
To reduce the risk of exploitation, we should therefore aim to discover vulnerabilities as soon as they are introduced into the code.
However, a previous study has shown that discovering bugs and vulnerabilities hidden deep in source code may require exponentially more executions~\cite{DBLP:conf/sigsoft/BohmeF20} compared to surface-level faults. Although continuous fuzzing may reduce the latency between code changes and bug detection, it does not lessen the expected time-to-detection and computational effort needed to uncover hard-to-reach faults.

To guide the allocation of additional fuzzing resources using vulnerability-risk estimates, we propose \APP (\app), a Just-In-Time approach that operates at the day level to match daily build cycles in large-scale fuzzing infrastructures (\eg OSS-Fuzz).
\APP estimates the risk of each project-day from fuzzing-based dynamic features in addition to the traditional code-change features used in previous studies.
While most prior JIT prediction work stops at predicting whether changes include bugs or vulnerabilities (\ie binary classification), \APP uses the estimated risk to decide whether and how much additional fuzzing should be performed.
The title borrows two moves from poker: for each project-day, \APP either \emph{checks}, keeping the regular fuzzing budget when the estimated risk is low, or \emph{raises}, allocating additional fuzzing sessions when the risk is high.
In this way, fuzzing effort is concentrated on high-risk changes and little is spent on low-risk ones, improving the cost efficiency of vulnerability detection in continuous fuzzing.

The main contributions of this research are as follows:
\begin{enumerate}
    \item \textbf{Just-In-Time Risk-Aware Resource Allocation for Continuous Fuzzing (RAISE).}
    We develop \APP, a JIT approach that uses vulnerability-risk estimates to determine whether and how much additional fuzzing should be performed in continuous fuzzing workflows.
    \smallskip
    \item 
    \textbf{Fuzzing-Derived Dynamic Features for Vulnerability-Risk Estimation.} 
    We introduce the use of fuzzing-derived dynamic features (\eg PatchCoverage) into vulnerability-risk estimation and prediction. To the best of our knowledge, prior work has not evaluated their use as predictive signals for vulnerability-introducing project-day changes and supplemental-fuzzing allocation in continuous-fuzzing environments. We therefore empirically evaluate their added value over established static features.
    \smallskip
    \item \textbf{Empirical Evaluation on a Large-scale Continuous-Fuzzing Dataset.}
    We construct a real-world dataset collected from OSS-Fuzz and the OSV, covering \finalprojects C/C++ projects, \finaltotaldays\xspace project-days, and \numberofvulnerabilities vulnerability reports, and use it to evaluate both vulnerability-risk estimation and additional fuzzing resource allocation in continuous fuzzing environments.
    \smallskip
    \item \textbf{Optimizing Resource Allocation Strategies for Additional Fuzzing.}
    We simulate five budget-constrained allocation strategies that schedule additional fuzzing on high-risk days.
    We show that the Simple-Regression strategy is significantly more cost-efficient than the prediction-agnostic heuristics, achieving the lowest pooled cost per vulnerability (\SfourBuildsPerDetection fuzzing sessions).
    \smallskip
    \item \textbf{Replication Package for Open Science.} 
    To support replication and further studies, we make our data and scripts available in a replication package on GitHub.\footnote{\url{https://github.com/kr37577/RAISE}}
\end{enumerate}

\smallskip\noindent
{\bf Structure of this paper:} Section \ref{sec:background} covers background and related work. Section \ref{sec:JIT} describes our methodology, and Section \ref{sec:Method} details the evaluation setup. Section \ref{sec:Result} presents the empirical results. Section \ref{sec:Threat} discusses threats to validity, and Section \ref{sec:Conclusion} concludes the paper.

\section{Background and Related Work}\label{sec:background} 
This section introduces continuous fuzzing and JIT defect/vulnerability prediction, and reviews related studies.

\begin{figure}[t]
    \centering
    \includegraphics[width=0.9\linewidth]{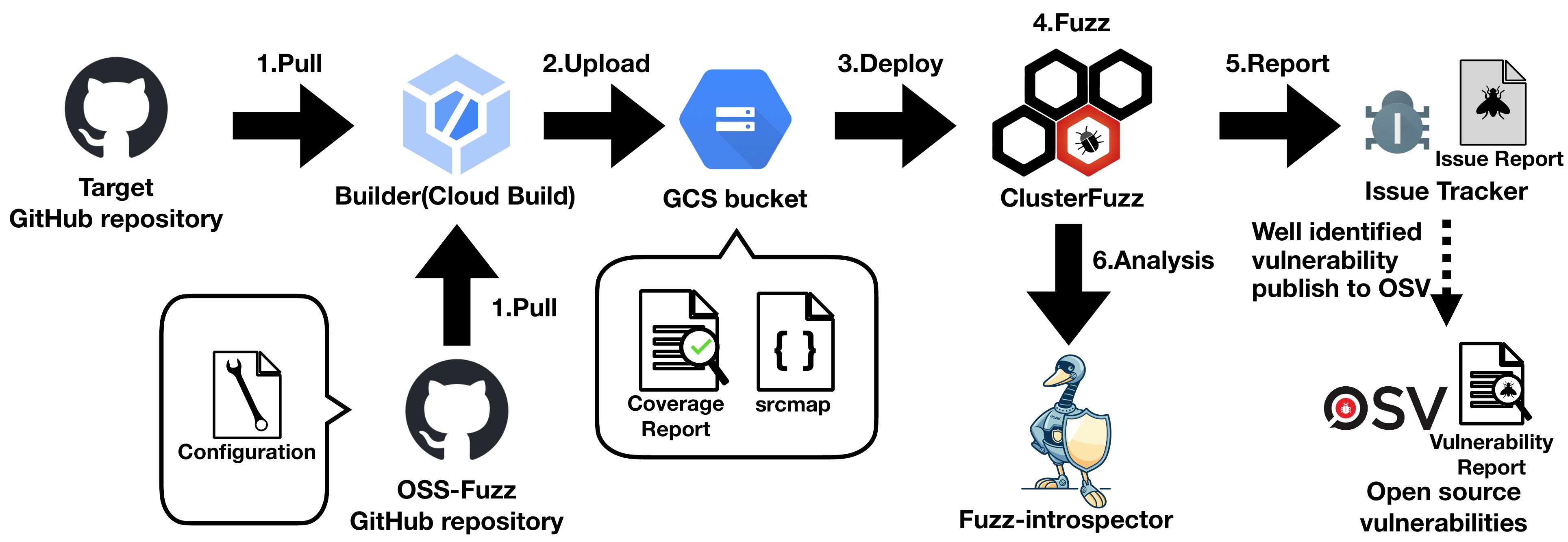}
    \caption{Overview of Fuzzing Process Using OSS-Fuzz
    }
    \label{fig:OSS-Fuzz-overview}
\end{figure}
\subsection{Fuzzing and Continuous Fuzzing}
In the early 1990s, Miller\etal evaluated 88 programs and found that over 24\% crashed when executed with random input strings~\cite{DBLP:journals/cacm/MillerFS90}. In that work, they introduced the concept of fuzzing, an automated approach for generating such inputs. Since then, fuzzing has evolved into a widely adopted technique, with several strategies proposed to improve both source code coverage and input generation. Despite its conceptual simplicity, fuzzing has proven highly effective in uncovering software vulnerabilities, often outperforming traditional static analysis approaches~\cite{DBLP:journals/compsec/CuiCHLDL22}.

Although fuzzing is effective for vulnerability detection, it often requires significant computational resources, involves long execution times~\cite{DBLP:conf/sigsoft/BohmeF20}, and poses challenges due to the complexity of setting up fuzzing environments~\cite{DBLP:journals/tosem/NourryKLBLK24}.
To alleviate these challenges, continuous fuzzing has emerged as a practical approach that integrates fuzzing into continuous integration (CI) workflows. This technique executes short fuzzing sessions either periodically or in batches within CI systems. 

One of the most popular implementations of continuous fuzzing is OSS-Fuzz. As of August 2025, OSS-Fuzz conducts large-scale continuous fuzzing on \currentTargetProject OSS projects. Figure~\ref{fig:OSS-Fuzz-overview} shows the overview of the fuzzing process with OSS-Fuzz. Based on the target OSS repository and the build configuration stored in the OSS-Fuzz repository, OSS-Fuzz builds each project by default once per day (Step 1).
The generated build artifacts are uploaded to cloud storage and utilized for tasks such as identifying regressions (Step 2). As part of this upload step, OSS-Fuzz also generates a metadata file, \texttt{srcmap.json}, for each build. This file maps source directories inside the build container to their corresponding version-control repositories (including the VCS type, repository URL, and commit hash for the main project and its dependencies), and is uploaded to the same cloud storage bucket so that downstream tools can reconstruct the exact project and dependency revisions used in a given fuzzing run.

The built projects are then fuzzed using ClusterFuzz, a tool for distributed fuzzing (Step 3, 4). In this process, not only are new fuzzing bugs detected, but previously identified bugs are also verified to ensure they no longer occur. When ClusterFuzz discovers a defect, an issue report is automatically generated and provided to developers (Step 5). The issue report includes information such as the ``Project Name'', ``Issue Number'', ``Detection Date and Time'', and ``Fix Date and Time'', in addition to ``Regressed range'' and ``Fixed range''. The Regressed range refers to a set/range of commits that could have introduced the vulnerability, while the Fixed range refers to a set/range of commits in which the fuzzing bug was likely fixed. 
The Regressed range is estimated using a binary search to identify which commits introduced the code where the vulnerability was detected. 
Specifically, ClusterFuzz repeatedly rebuilds and runs the minimized reproducer on midpoint revisions between a known non-crashing and crashing revision, halving the interval until it establishes the crashing window.
The fixed range is determined similarly by bisecting between a crashing revision and a later non-crashing revision to locate where the reproducer stops crashing.

Once a fix is confirmed or 90 days have passed since detection, the fuzzing bug is made public by creating an issue report in the official OSS-Fuzz issue tracker, and the corresponding report is updated. 
Security bugs detected by OSS-Fuzz are automatically ingested and published as vulnerability entries in the OSV database following a defined disclosure process.
In addition, OSS-Fuzz generates coverage reports at the end of each fuzzing session. 
Fuzz Introspector analyzes these reports by combining dynamic coverage data with static reachability information, after which the reports are published to provide insights such as line, function, and region coverage (Step 6).




\subsection{Just-In-Time Defect and Vulnerability Prediction}
Defect prediction aims to estimate which parts of a software system are likely to contain defects. Just-In-Time (JIT) defect prediction focuses specifically on code changes and predicts whether a newly committed change is likely to introduce a defect. By making predictions at the time of change, JIT defect prediction can support quality assurance activities by identifying changes that may require additional review or testing.

A representative JIT defect prediction model was proposed by Kamei et al.~\cite{DBLP:journals/tse/KameiSAHMSU13}. Their model uses 14 metrics organized into five dimensions: diffusion, size, purpose, history, and developer experience. For example, these metrics capture how widely a change is distributed across files, how many lines are added or deleted, whether the change fixes a defect, and the experience of the developer who made the change. These metrics characterize the risk of a change primarily from its modifications and development history.

JIT prediction has also been applied to security-related code changes. VCCFinder~\cite{DBLP:conf/ccs/PerlD0AYRFA15}, for example, predicts vulnerability-contributing commits using not only repository metadata but also information extracted from the contents of code changes. Its features include characteristics such as the number of modified functions, C/C++ keywords appearing in a patch, and textual information from the patch itself. VCCFinder therefore extends change-risk prediction with information about what was actually modified in the source code.

This study uses these JIT prediction metrics to estimate vulnerability risk and uses it to determine whether and how much additional fuzzing should be allocated during continuous fuzzing.

\subsection{Related Work}
\label{sec:related}
\subsubsection{Continuous Fuzzing}

With the release of OSS-Fuzz, empirical studies on fuzzing and continuous fuzzing have become increasingly popular.
Ding\etal\cite{DBLP:conf/msr/DingG21} analyzed OSS-Fuzz issue reports to investigate the characteristics of detected bugs and their lifecycle.
Their findings showed that OSS-Fuzz discovers most regressions within one week.
Keller\etal\cite{DBLP:conf/msr/KellerMM23} also investigated the lifecycle of bugs using issue reports from OSS-Fuzz. Their analysis revealed that while the median time from detection to fix was relatively short (around 2 days), the median time from bug introduction to fix was substantially longer, with a median of 324 days.
Zhu\etal\cite{DBLP:conf/ccs/ZhuB21} investigated the characteristics of bugs detected by OSS-Fuzz and found that 77\% of the bugs were recently introduced. This indicates that continuous fuzzing is particularly effective at identifying newly introduced faults.

Beyond bug detection, researchers have also examined the infrastructure of continuous fuzzing.
G{\"{o}}rz\etal\cite{DBLP:journals/corr/abs-2505-06177} conducted a large-scale analysis of harness degradation in OSS-Fuzz using harness update information and coverage data. Their study revealed that the average coverage improvement remains marginal at only 0.26\%.
Nourry\etal\cite{DBLP:journals/tosem/NourryKSSK25} analyzed OSS-Fuzz build logs to investigate the nature of build failures in fuzzing. Their study revealed that approximately 9.41\% of builds fail.
Our work differs by using continuous fuzzing data to detect vulnerabilities in source code changes.

\subsubsection{Vulnerability Prediction}
Extensive research has been conducted on software vulnerability detection.
Russell\etal\cite{DBLP:conf/icmla/RussellKHLHOEM18} applied deep representation learning to OSS functions. They showed that vulnerability detection achieves high performance on both the SATE IV benchmark and real-world code by combining CNNs/RNNs with random forests.
Zhou\etal\cite{DBLP:conf/nips/ZhouLSD019} focused on capturing semantic features of code structure for function-level vulnerability detection and proposed Devign.
By combining a Gated Graph Recurrent Network with a convolutional module, their proposed model achieves over 10\% higher accuracy and F1 score than previous methods.
Fu\etal\cite{DBLP:conf/msr/FuT22} proposed LineVul, a Transformer-based approach leveraging CodeBERT. They achieved a 160–379\% improvement over existing methods in function-level vulnerability detection and realized a 12–25\% improvement in Top-10 accuracy for line-level localization.
Yildiz\etal\cite{DBLP:conf/acl/YildizTLF0D25} constructed the JITVUL benchmark to evaluate vulnerability detection under practical conditions. Their comparison of LLMs and ReAct agents revealed that relying solely on F1 score can lead to overestimation, highlighting the importance of incorporating pairwise precision.

Furthermore, extensive work on Just-In-Time (JIT) defect prediction has proposed commit-level models for identifying risky changes. 
Hoang\etal\cite{DBLP:conf/msr/HoangDK0U19} developed DeepJIT, which predicts defects by using both commit messages and code changes. Their evaluation showed that DeepJIT achieved up to 13\% improvement over the state-of-the-art model, highlighting the critical role of commit messages in defect prediction.
Guglielmi\etal\cite{DBLP:conf/msr/GuglielmiDOS25} investigated incorporating developer-centric features into JIT defect prediction models and showed that such features improve prediction performance.
Recent studies have also explored vulnerability detection using JIT approaches.
Lomio\etal\cite{DBLP:journals/jss/LomioILPL22} investigated how effectively JIT bug prediction can be applied to vulnerability prediction. They concluded that vulnerability prediction based on commit-level bug prediction is not yet practical and emphasized the importance of investigating features for identifying vulnerabilities.
Perl\etal\cite{DBLP:conf/ccs/PerlD0AYRFA15} proposed VCCFinder, which detects risky commits by integrating code-difference features with development process metadata. Their results showed that leveraging both code features and metadata is effective in improving detection performance.
Nguyen\etal\cite{DBLP:journals/jss/NguyenNVDNV24} focused on the semantics of code changes and proposed CODEJIT, a framework for vulnerability detection that considers not only modified code but also its impact on unmodified segments.
Nguyen\etal\cite{DBLP:journals/corr/abs-2507-10729} found that evaluating on vulnerability-introducing or -fixing commits inflates performance, with existing methods degrading substantially when non-vulnerability-inducing commits are included.

\subsubsection{Prediction-Guided Fuzzing}

Several studies have incorporated prediction or risk estimation into fuzzing to improve its efficiency.
Zong\etal\cite{DBLP:conf/uss/ZongLWDL020} proposed FuzzGuard, which predicts whether generated inputs can reach target code and filters out unreachable inputs before execution. Their evaluation on 45 real vulnerabilities showed that FuzzGuard achieved a speedup of up to 17.1$\times$ over AFLGo.
Lipp\etal\cite{DBLP:conf/issta/LippEKPBB23} proposed Green Fuzzing, which uses vulnerability prediction and coverage saturation to determine when a fuzzing campaign should be stopped. Their evaluation showed that 24-hour fuzzing campaigns could be terminated 6--12 hours earlier while missing fewer than 0.5 bugs on average.
Cui\etal\cite{DBLP:journals/scp/CuiJCWL24} proposed DPFuzz, which predicts defect-prone modules and allocates more fuzzing resources to test cases that exercise those modules. In their evaluation on mJS, DPFuzz triggered 26.1\% more crashes on average than FairFuzz.
Lee\etal\cite{DBLP:conf/asiaccs/LeeXSLP24} proposed SyzRisk, a continuous kernel regression fuzzer that estimates the risk of recent code changes based on recurring change patterns and prioritizes fuzzing toward risky changes. Their evaluation on Linux kernel v6.0 showed a 61\% improvement in bug exposure time compared with Syzkaller.

Prior work uses coverage to guide input generation, prioritize code regions, or stop fuzzing campaigns. In contrast, \APP uses coverage observations from regular continuous fuzzing as predictive features to estimate whether recently introduced changes are vulnerability-prone and to determine whether supplemental fuzzing should be scheduled.

Our work differs from prior studies in two key aspects. First, while existing vulnerability prediction methods are designed for general software development contexts, \APP is tailored specifically to continuous fuzzing environments. \APP incorporates execution feedback from regular fuzzing, such as coverage metrics, into vulnerability-risk estimation and predicts vulnerability-inducing changes on a daily basis to match the daily fuzzing cycle of systems like OSS-Fuzz. Second, while prior prediction-guided fuzzing studies use prediction or risk information to prioritize inputs, target code regions, or to determine when a fuzzing campaign should be stopped, \APP addresses a different decision: whether, and to what extent, additional fuzzing sessions should be performed after regular continuous fuzzing. This enables more efficient use of computational resources in large-scale OSS fuzzing environments.

\begin{figure*}[!t]
    \centering
    \includegraphics[width=0.95\linewidth]{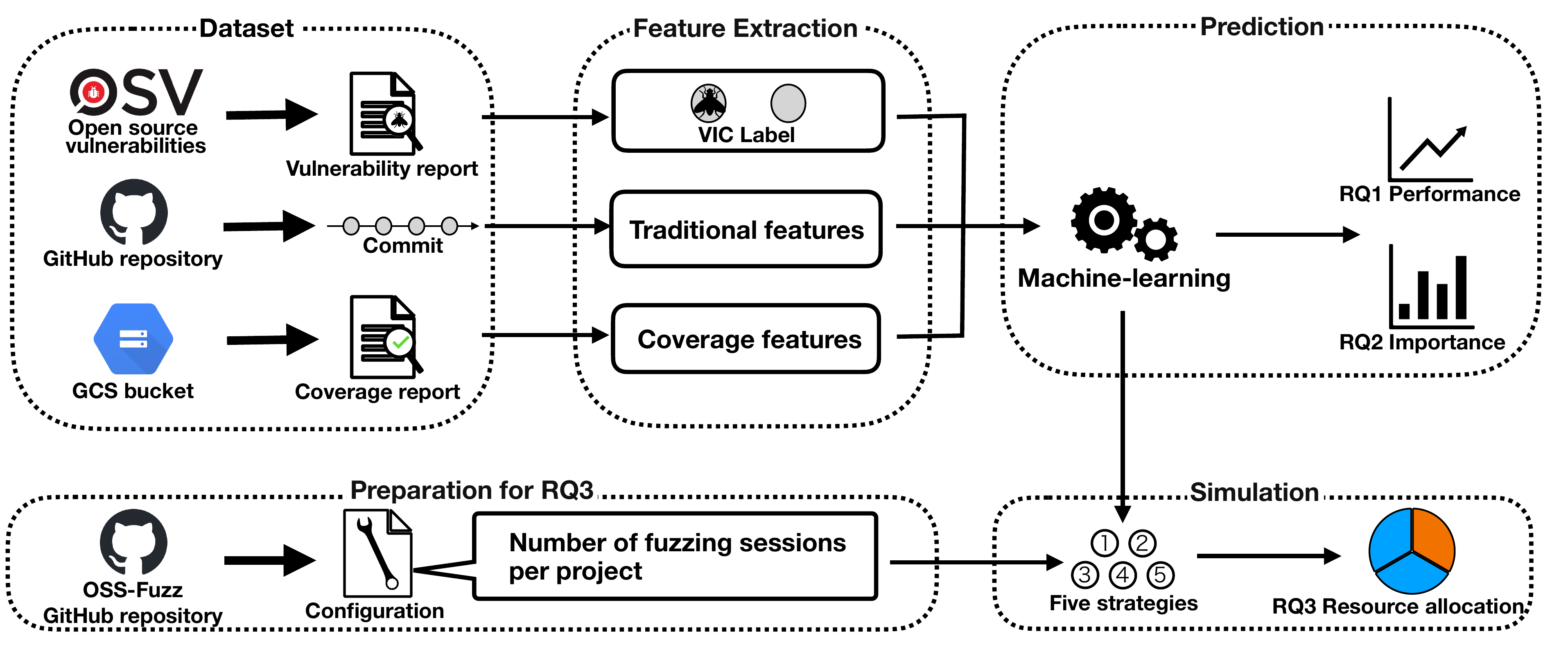}
    \caption{Overview of \app 
    }
    \label{fig:overview}
\end{figure*}
\section{\APP: \app}\label{sec:JIT}

\subsection{Overview}\label{sec:overview} 
We propose \textit{\app}~(\APP), which estimates the vulnerability risk of code changes and adaptively allocates additional fuzzing to high-risk changes.
The model guides decisions to maintain standard fuzzing or perform additional fuzzing when vulnerability-prone changes occur.
\autoref{fig:overview} illustrates the overall framework of our approach.
\APP~utilizes features extracted from various software processes, target software systems, and fuzzing.
These features are preprocessed and fed into a machine learning model to solve a binary classification problem: whether the provided code changes include at least one vulnerability or not.
Once \APP{} detects vulnerability-prone changes, one of five strategies can be applied to increase the amount of fuzzing performed. 

\subsection{Vulnerability-prone Change Detection}

\subsubsection{Feature extraction}\label{sec:feature_extraction} 
We build \APP using an \allfeature-dimensional feature set that combines established static code metrics with fuzzing-derived dynamic features, focusing on commits modifying C/C++ source files.

\noindent
\textbf{Baseline Features:}\label{sec:baseline_features}
Given a code change, we characterize its static properties by employing \tworesearchfeature features from two previous studies on JIT defect and vulnerability prediction.
\begin{itemize}
    \item \textit{Change Features}: Kamei\etal\cite{DBLP:journals/tse/KameiSAHMSU13} developed a set of 14 process and product quality metrics. These features measure the size of the change (\eg lines added/deleted) and the experience of the developers making the change.
    Developer-experience metrics are automatically extracted from each project's Git history using the commit author's e-mail and prior commit history.

    \item \textit{C/C++ Syntax-based Change Features}: Using parts of Riom\etal's publicly available replication package \cite{DBLP:journals/ese/RiomSABMK21} for VCCFinder proposed by Perl\etal\cite{DBLP:conf/ccs/PerlD0AYRFA15}, we extracted the available VCCFinder-based features. We excluded the four ``Expressions'' metrics (F30--F33) because they were not implemented in the released feature extractor. Consequently, the impact of excluding these four metrics on predictive performance was not evaluated in this study.
    
\end{itemize}

\noindent
\textbf{Fuzzing-based Dynamic Features:} \label{sec:fuzzing_features} 
To enhance the performance of JIT vulnerability prediction, we introduce dynamic features derived from fuzzing activities.
Although fuzzing can uncover many vulnerabilities, it often fails to achieve full coverage of the target software system or of the code changes made by a commit.
Therefore, the extent of achieved coverage provides a critical signal regarding the reliability of the code.
To capture this information, we leverage code coverage reports published by OSS-Fuzz \cite{203944}. Using the commit date, we associate each commit with its corresponding coverage report. As shown in \autoref{tab:proposed_fuzzing_features}, we extract \typeofcoveragefeature types of coverage metrics across several levels of granularity: line, function, region, branch, instantiation, and patch coverage. 
Because CoverageDelta is computed for the five project-level metrics (i.e., all metrics except PatchCoverage), these feature types yield 11 dynamic features in total. Combined with the \tworesearchfeature static features (14 Kamei\etal features and 63 VCCFinder-based features), they form the \allfeature-dimensional feature set.

\begin{table*}[t] 
\footnotesize
    \centering
    \caption{Fuzzing-based Dynamic Features}
    \label{tab:proposed_fuzzing_features}
    \begin{tabular}{@{}lp{0.25\textwidth}p{0.5\textwidth}@{}}
        \toprule
        \textbf{Feature Name} & \textbf{Definition}                                                                                                                                                                                & \textbf{Rationale}                                                                                                                                                \\
        \midrule
        LineCoverage          & The percentage of executable code lines that have been executed at least once~\cite{clangdocs}.                                                                                                                     & Lines must be executed for vulnerabilities to be found; commonly used to compare fuzzer effectiveness~\cite{DBLP:conf/sp/ChenC18}.                                 \\
        \addlinespace
        FunctionCoverage      & The percentage of functions that have been executed at least once~\cite{clangdocs}.                                                                                                                                 & A function must be executed during fuzzing to detect its vulnerabilities.                                                                                                             \\
        \addlinespace
        RegionCoverage        & The percentage of code regions that have been executed at least once~\cite{clangdocs}.                                                       & 
        Code region denotes a compiler generated coverage mapping unit~\cite{llvmdocs}. Region coverage may offer finer granularity than line coverage and can therefore assist in tracking day-to-day fuzzing progress and in revealing newly untested code.
        \\
        \addlinespace
        BranchCoverage        & The percentage of conditional branches that have been executed at least once~\cite{clangdocs}.                                                                                                                                  & Fuzzer evaluations have reported a very strong correlation between branch coverage and bug discovery~\cite{DBLP:conf/icse/BohmeSM22}.                             \\
        \addlinespace
        InstantiationCoverage & The percentage of function template instances that were executed at least once~\cite{clangdocs}.                                                                                                                  &
        C++ function templates are instantiated as multiple functions at compile time, depending on the type arguments. Function coverage does not distinguish between these concrete instantiations. InstantiationCoverage reveals whether there are any unexecuted instantiations of functions remaining.
        \\
        \addlinespace
        PatchCoverage         & The percentage of lines added or modified in the current commit's patch that were executed at least once.                                                                                          &  Targeting patch regions improves regression detection efficiency \cite{DBLP:conf/ccs/BohmePNR17}. 
        Patch coverage shows how much newly modified code is executed, which in continuous fuzzing is more directly tied to vulnerability detection than existing coverage~\cite{DBLP:journals/tse/ShiraiNKFKI26}. 
                                                                                                                                                              \\
        \addlinespace
        CoverageDelta         & The change in coverage compared to the previous day (e.g., $Coverage(day_{i}) - Coverage(day_{i-1})$). 
        & CoverageDelta is used to detect sudden drops or increases in testing coverage. Day-to-day changes in coverage reveal test progress or degradation; sudden drops may indicate untested new code, while steady gains reflect effective exploration.  \\
        \bottomrule
    \end{tabular}
\end{table*}

\subsubsection{Label Creation}\label{sec:lable_creation}

To identify vulnerability-introducing commits (VICs), we leverage the Open Source Vulnerabilities (OSV) database, which catalogs known vulnerabilities in open-source software~\cite{OSV}.
Specifically, we extract the SHAs of VICs from the ``introduced'' fields of OSV entries. These commits are labeled as VICs.
In our classification setup, a positive label denotes that the commit introduces at least one vulnerability, while a negative label indicates the absence of any known vulnerabilities. 




\subsection{Strategies for Additional Fuzzing}
\label{sec:resource_allocation} 
This section defines five strategies to increase the amount of fuzzing when \APP predicts that a set of commits/changes committed on a given day contains vulnerabilities. Note that, for all strategies, \APP recommends increases in terms of additional fuzzing sessions to conduct (rounded to the nearest integer).
To avoid information leakage, all strategies are parameterized in a leave-one-project-out manner: the statistics and regression models used for a target project are derived only from the other projects.
\begin{enumerate}
\item{\textbf{Median Strategy:}}
This strategy suggests additional fuzzing sessions based on the median number of fuzzing sessions that were required to find past vulnerabilities in projects participating in OSS-Fuzz.
\smallskip
\item{\textbf{Randomized Strategy:}}
This strategy suggests additional fuzzing sessions by assigning a random number between the first and third quartiles when calculating the required number of fuzzing sessions that were required to identify past vulnerabilities across the other projects.
\smallskip
\item{\textbf{Line-Change Proportional Strategy:}}
Inspired by Monden\etal~\cite{DBLP:journals/tse/MondenHSSYBM13}, this strategy allocates additional fuzzing sessions proportionally to the size of changes. 
The number of additional fuzzing sessions is calculated as follows:
\[
a_i =  F_{median} \cdot \frac{\log(D_i + 1)}{\log(D_{median} + 1)}
\]
\smallskip
where $a_i$ represents the number of additional sessions allocated to day $i$. 
$F_{median}$ is the median number of additional fuzzing sessions that were required to find past vulnerabilities in projects participating in OSS-Fuzz. 
$D_i$ is the size of changes (measured in lines of code modified) on day $i$, and $D_{median}$ is the median change size within the same project. 
This proportional allocation ensures that days with larger code changes undergo more fuzzing, reflecting the assumption that larger modifications have a higher potential for introducing vulnerabilities. 


\smallskip
\item{\textbf{Simple-Regression Strategy:}}
This strategy predicts the number of additional fuzzing sessions required using a simple regression model. The simple linear regression model is trained using the number of modified source code lines on a given day and the number of fuzzing sessions required to find previous vulnerabilities.

\smallskip
\item{\textbf{Multi-Regression Strategy:}}
This strategy predicts the number of additional fuzzing sessions required using a multiple regression model. The multiple linear regression model is trained using all the metrics used for \APP (See Section~\ref{sec:feature_extraction}) and historical data (\ie the number of sessions that were required to find previous vulnerabilities).

\end{enumerate}

\section{Experimental Setup}
\label{sec:Method} This section details the approach and experimental design
adopted to answer our research questions. 

\subsection{Research Questions}
\label{sec:rq}
We address three research questions to evaluate the feasibility and practical utility of \APP. 

\medskip
\noindent\textbf{\textit{\rqA}}\par
Several studies have attempted to identify commits that introduce defects or vulnerabilities, an approach known as JIT bug prediction.
Kamei\etal\cite{DBLP:journals/tse/KameiSAHMSU13} proposed an approach to detect defect-inducing commits using 14 metrics extracted from version control systems, while Perl\etal\cite{DBLP:conf/ccs/PerlD0AYRFA15} predicted vulnerability-inducing commits using code diffs and commit messages.
In the context of continuous fuzzing, Zhu\etal\cite{DBLP:conf/ccs/ZhuB21} reported that 77\% of 23,000 bugs discovered by OSS-Fuzz were regressions introduced by recent code changes. This finding indicates that recently introduced changes account for a substantial fraction of bugs discovered during continuous fuzzing. However, existing JIT prediction models have mainly been evaluated using defects or vulnerabilities collected from general software-development data. It therefore remains unclear how effectively these models can identify vulnerability-introducing changes that are subsequently discovered through continuous fuzzing.
Therefore, in RQ1, we investigate whether established JIT prediction techniques can identify vulnerability-introducing project-days in a continuous fuzzing environment.

\medskip
\noindent\textbf{\textit{\rqB}}\par

Prior JIT defect and vulnerability prediction studies primarily rely on static metrics derived from code changes and development history~\cite{DBLP:journals/tse/KameiSAHMSU13,DBLP:conf/ccs/PerlD0AYRFA15}. In continuous fuzzing, however, regular fuzzing performed after code changes provides execution feedback, such as code coverage. Because fuzzing explores program behavior through dynamically generated inputs and its exploration progress can vary across executions, code-change and development-process metrics alone do not reveal how thoroughly the current revision, particularly the modified code, has actually been exercised. Prior large-scale empirical work on continuous fuzzing has shown that changes in code coverage are associated with fuzzing bug discovery~\cite{DBLP:journals/tse/ShiraiNKFKI26}.
This suggests that execution feedback from continuous fuzzing may provide useful information for identifying vulnerability-prone changes.
In RQ2, we investigate whether incorporating fuzzing-derived execution feedback improves vulnerability prediction beyond established static metrics, and which metrics contribute most to the prediction.


\smallskip
\noindent\textbf{\textit{\rqC}}\par
Prior studies have used defect or vulnerability predictions to improve fuzzing resource decisions. For example, DPFuzz allocates more testing resources to modules predicted to be defect-prone, while Green Fuzzing uses predicted vulnerable functions to determine when a fuzzing campaign should be stopped~\cite{DBLP:journals/scp/CuiJCWL24,DBLP:conf/issta/LippEKPBB23}. However, these studies focus on how resources are used during a fuzzing campaign; they do not address whether additional fuzzing should be performed after regular continuous fuzzing nor how much additional effort should be allocated. This decision is important because predicting vulnerability-prone changes does not guarantee that regular fuzzing will expose the corresponding vulnerabilities and indiscriminately increasing fuzzing efforts would incur unnecessary computational cost.
In RQ3, we investigate how vulnerability-risk estimates can be leveraged to dedicate additional fuzzing effort. Specifically, we compare five allocation strategies and evaluate them based on their efficiency at discovering vulnerabilities.


\subsection{Data Collection}
\label{sec:dataset} 

\subsubsection{Vulnerability Labeling}
To train models, we need to create labels indicating whether a set of commits made on a given day $i$ has at least one vulnerability. To do so, we utilize the OSV database~\cite{OSV}, which compiles reports of vulnerabilities discovered in the wild. The database includes the project name, the publication timestamp of the vulnerability entry, aliases (\eg CVE and GitHub Security Advisory (GHSA) identifiers), affected versions and commit ranges (including the introduced and fixed commits), the repository URL, and references to advisories, bug reports, and fixing commits. In total, we collected \allvulnerabilityreport OSV vulnerability reports from the database.  

The publication timestamp in OSV indicates when an entry was published or updated, not when the vulnerability was actually detected. To obtain the detection date and detailed information, we refer to the OSS-Fuzz issue reports associated with each OSV entry. OSS-Fuzz generates an issue report upon discovering a vulnerability, and each report contains the stack trace, a reproducible test case, regression ranges, and metadata such as the fuzzing engine, sanitizers used, fuzz target name, and platform. The issue description also includes a reported timestamp, which we use as the vulnerability detection time, following the same approach as prior work~\cite{DBLP:conf/msr/DingG21}. 
All \allvulnerabilityreport OSV entries included at least one reference to the corresponding OSS-Fuzz issue report.

\subsubsection{Project Selection}
To compare our results with previous work that supports only C/C++ features (\eg VCCFinder~\cite{DBLP:conf/ccs/PerlD0AYRFA15}), we restricted our data collection to OSV vulnerability reports that are linked to OSS-Fuzz issue reports and whose repositories are primarily written in C/C++. 
This resulted in \ccppvulnerabilities OSV vulnerability reports (\ccpppercent\% of total reports) from \ccppprojects repositories. 
After filtering out repositories that were not publicly available and could not be cloned, we were left with \clonedvulnerabilities OSV vulnerability reports from \clonedprojects repositories.

Because \APP makes daily predictions, we need projects with sufficient historical data for reliable predictions. We therefore filtered out projects with fewer than 10 vulnerability introducing days. As a result, our dataset consists of \finalprojects projects and \finalpositivedays vulnerability introducing days, derived from \numberofvulnerabilities distinct OSV vulnerability reports.

\smallskip
\subsubsection{Coverage Data Collection}
Because our model leverages code coverage metrics from fuzzing, we also collected daily coverage reports archived in the public OSS-Fuzz Google Cloud Storage (GCS) bucket. 
Our coverage report dataset consists of \finalcoveragereport coverage reports collected from the aforementioned \finalprojects projects dating from \scrapestartdate to \scrapeenddate. These coverage reports contain project wide coverage metrics for each day, such as executed lines, functions, regions, branches, and instantiation coverage.

To compute the patch coverage defined in Table \ref{tab:proposed_fuzzing_features}, we mined \finalsrcmap srcmap.json files generated during our studied period. A srcmap.json file maps source directories in the build container to their version-control metadata (\eg repository URL and commit hash). 
We parse each srcmap.json file generated by the daily coverage builds to identify the exact project revisions that were built.
After identifying the exact revisions, we then extract the source code diffs between each revision and its parent.
Finally, we measure the daily PatchCoverage by calculating the percentage of added/modified lines from the diff that were executed at least once.
Note that the srcmap.json file is used to identify the exact revision built for coverage reporting, not to infer when a vulnerability was introduced.
\autoref{tab:dataset_stats} summarizes the  statistics of the final dataset.
\begin{table}[ht]
    \centering
    \caption{Statistics of the Studied Dataset}
    \label{tab:dataset_stats}
    \begin{tabular}{lr}
        \toprule \textbf{Metric}        & \textbf{Value}     \\
        \midrule Number of Projects     & \finalprojects     \\
        Total Days (Instances)          & \finaltotaldays    \\
        \# Positive Days (with VICs)    & \finalpositivedays \\
        \# Negative Days (without VICs) & \finalnegativedays \\
        Ratio of Positive Days          & \finaldaysratio    \\
        Number of coverage report       & \finalcoveragereport \\
        Number of srcmap                & \finalsrcmap \\
        Number of vulnerability report  & \numberofvulnerabilities \\
        \bottomrule
    \end{tabular}
\end{table}

\begin{table}[t]
    \footnotesize
    \centering
    \caption{Daily Feature Aggregation Rules}
    \label{tab:aggregation_rules}
    \begin{tabular}{p{0.1\linewidth} p{0.05\linewidth} p{0.7\linewidth}}
        \toprule \textbf{Category}              & \textbf{Rule} & \textbf{Rationale / Hypothesis}                                                                                                                                                                                                                                     \\
        \hline
        \addlinespace Kamei \& VCCFinder& \texttt{sum}              & For size-based and count-based Kamei metrics and all VCCFinder features, \emph{sum} aggregates the total magnitude of code modifications and the total count of potentially risky patterns introduced throughout the day. \\
        \hline
        \addlinespace Kamei & \texttt{mean}             & For developer experience and entropy-related Kamei metrics, \emph{mean} provides the average characteristic of the day's commits, smoothing out the influence of any single outlier commit. \\
        \hline
        \addlinespace Coverage          & \texttt{last}             & Metrics that represent a final state capture the definitive state of test coverage at the end of the day. This value provides the most appropriate baseline for subsequent development.    \\
        \bottomrule
    \end{tabular}
\end{table}

\subsection{Feature Extraction} \label{sec:delta_features}
In this experiment, we predict whether or not the set of commits made on a given day $i$ contains at least one vulnerability. While most JIT defect prediction models operate at the individual commit level, our approach aggregates commits on a daily basis. This design choice is motivated by the operational characteristics of OSS-Fuzz, which evaluates coverage daily rather than per commit, making it impractical to compute accurate coverage-related metrics for individual commits. 
We therefore aggregate commit-level metrics into daily-level metrics. We excluded days with no commits and days for which no OSS-Fuzz coverage report was generated. The aggregation logic varies depending on the feature type in order to best capture the characteristics of the changes made on a given day. \autoref{tab:aggregation_rules} summarizes the transformation rules applied to each feature type.

Similarly, we assign a binary label (1: includes at least one vulnerability; 0: includes none) for any given day $i$ based on whether the set of commits made on that day is known to contain at least one vulnerability.


\subsection{Model Construction}\label{sec:model_construction} 

We employ two machine learning algorithms that have demonstrated strong performance in JIT defect prediction \cite{DBLP:journals/sqj/DinterCGT23,DBLP:journals/ese/WimalasooriyaLCM24,DBLP:conf/msr/FukushimaKMYU14,DBLP:journals/ese/ZhouHL25}, Random Forest~\cite{DBLP:journals/ml/Breiman01} and XGBoost~\cite{DBLP:conf/kdd/ChenG16}. These algorithms are known for their effectiveness with limited data and robustness against overfitting~\cite{DBLP:journals/ml/Breiman01,DBLP:conf/kdd/ChenG16,DBLP:conf/nips/GrinsztajnOV22}. Moreover, they provide interpretable feature-importance scores, supporting our goal of improving our understanding of which features are effective for vulnerability prediction.

We use fixed hyperparameter settings for Random Forest and XGBoost across all projects and feature configurations, without hyperparameter tuning. By intentionally holding model settings constant, we isolate the effect of the feature configurations and prevent differences in tuning from confounding their comparison.

Since only \finaldaysratio of the samples belong to the positive class in our dataset, we applied random undersampling to address class imbalance during model training. Specifically, we randomly selected a subset of the majority class (\ie negative class) to match the size of the minority class (\ie positive class), thereby balancing the training data.

\subsection{Evaluation}
\label{sec:evaluation}
To simulate a realistic deployment scenario in which a model is applied to a new project that lacks historical labeled data, we adopt a cross-project setting~\cite{DBLP:journals/ese/KameiFMYUH16}. Unlike within-project, cross-project prediction models are trained on historical data from other projects to predict vulnerabilities in new, unseen target projects.
Given a set of $N$ projects, we iteratively select one project as the target and use the remaining $N-1$ projects as the training set. 
For the testing phase, we strictly preserve the chronological order of days within the target project.
This approach ensures that the evaluation respects temporal dependencies, which is critical for time-sensitive evaluation settings, as emphasized by recent research~\cite{DBLP:journals/csur/ZhaoDC23}.

Furthermore, to mitigate the impact of random variations inherent to machine learning algorithms (\eg initialization seeds in RF and XGBoost) and data sampling, we repeat the training and evaluation procedure 30 times with different random seeds for each target project and report the mean performance. This evaluation protocol follows recent work on JIT defect prediction~\cite{DBLP:conf/wcre/MoritaKKSMKU24}.



While the prediction model outputs a probability \(p \in [0,1]\) indicating whether a given instance contains a vulnerability, we convert these probabilistic outputs into binary labels using a default decision threshold of \(p = 0.5\), as is commonly done in defect and vulnerability prediction studies~\cite{DBLP:journals/tse/KameiSAHMSU13,DBLP:conf/msr/FuT22}. 
Following this practice, instances with \(p \ge 0.5\) are classified as positive when computing threshold-dependent evaluation metrics. Under this setting, \APP~performs a binary classification task to determine whether the set of code changes made on a given day contains at least one vulnerability-introducing change.

Given the class imbalance in our dataset, we adopt ROC-AUC, PR-AUC, and Matthews Correlation Coefficient (MCC) as our primary performance measures, rather than traditional metrics such as precision and recall. 
ROC-AUC and PR-AUC are threshold-independent ranking-based measures, and MCC is a complementary confusion-matrix-based statistic that is often recommended for imbalanced defect/vulnerability prediction.
These metrics are better suited for evaluating models under imbalanced conditions, as demonstrated in recent studies~\cite{DBLP:journals/corr/abs-2507-10729,DBLP:journals/tse/Tantithamthavorn20,DBLP:journals/tosem/CaoSBWLWTZL24}. For completeness, we also report traditional metrics as a reference. 
The detailed definitions for each performance measure are described below.

\smallskip
\noindent
\textbf{Precision, Recall, and F1-Score}: These metrics provide insight into the trade-offs of the model. Precision measures the correctness of positive predictions, which is critical for minimizing developer effort spent on false alarms. Recall (or sensitivity) measures the model's ability to find actual positive instances, which is paramount for security. The F1-Score provides the harmonic mean of Precision and Recall, offering a balanced view of performance.

\smallskip
\noindent
\textbf{PR-AUC}: PR-AUC (Area Under the Precision-Recall Curve) summarizes the trade-off between precision and recall across all classification thresholds. Unlike ROC-AUC, PR-AUC focuses on the performance of the minority class, making it highly informative for imbalanced datasets.

\smallskip
\noindent
\textbf{ROC-AUC}: ROC-AUC (Area Under the ROC Curve) is a metric that measures the model's ability to distinguish between positive and negative classes across all thresholds. It is worth noting that it can be overly optimistic on imbalanced data due to the large number of true negatives \cite{DBLP:journals/ese/ZhouHL25}.

\smallskip
\noindent
\textbf{MCC}: MCC (Matthews Correlation Coefficient) is a correlation coefficient between the observed and predicted classifications, calculated from all four entries of the confusion matrix. It is widely regarded as one of the most reliable single-score metrics for imbalanced binary classification.

	\smallskip
\noindent
\textbf{Statistical analysis}: To determine whether the observed differences are statistically significant rather than due to chance, we treat each of the 15 projects as a single paired observation and compare configurations using the two-sided Wilcoxon signed-rank test~\cite{Wilcoxon}. We choose this non-parametric, paired test because our samples are small ($N{=}15$), not normally distributed, and naturally paired by project (the same project is evaluated under every configuration). For each project, the 30 repetitions and cross-validation folds are not independent observations; they repeatedly remeasure the same underlying project data under different random seeds. We therefore average them into a single value per project before testing because treating each repetition/fold outcome as an independent sample would otherwise inflate the effective sample size (\ie pseudo-replication) and make the test appear more significant than it really is. We complement each test with the Cliff's~$\delta$ effect size~\cite{Cliff}, interpreting $|\delta|$ as negligible ($<0.147$), small ($<0.33$), medium ($<0.474$), or large otherwise. Because we perform many pairwise comparisons, we control the false discovery rate using the Benjamini--Hochberg procedure~\cite{Benjamini} and report BH-adjusted $p$-values, considering a difference statistically significant when the adjusted $p<0.05$. We apply this same protocol uniformly across all three RQs. In RQ3, each project contributes one cost-per-vulnerability value per strategy, and projects in which either strategy of a pair detected no vulnerability are excluded from that pair;
strategies are therefore compared with the same project-paired test.

\begin{table*}[t]
    \centering
    \caption{Ablation Results: Daily Vulnerability Prediction Performance Across Six Feature Sets}
    \label{tab:full_results}
    \resizebox{\textwidth}{!}{%
    \begin{tabular}{@{}lllrrrrrr@{}}
        \toprule
        Feature Set ID & Feature Set & Model & ROC-AUC & PR-AUC & MCC & Precision & Recall & F1-Score \\
        \midrule
        \multirow{1}{*}{fs0} & \multirow{1}{*}{Random} & --- &
        \expZeroRandomRocAucValue & \expZeroRandomPrAucValue & \expZeroRandomMccValue &
        \expZeroRandomPrecisionValue & \expZeroRandomRecallValue & \expZeroRandomFoneValue \\
        \midrule
        \multirow{2}{*}{fs1} & \multirow{2}{*}{Kamei} & RF  &
        \expOneRfRocAucValuenum & \expOneRfPrAucValuenum & \expOneRfMccValuenum &
        \expOneRfPrecisionValuenum & \expOneRfRecallValuenum & \expOneRfFoneValuenum \\
        & & XGB &
        \expOneXgbRocAucValuenum &  \expOneXgbPrAucValuenum & \expOneXgbMccValuenum & \expOneXgbPrecisionValuenum & \expOneXgbRecallValuenum & \expOneXgbFoneValuenum \\
        \midrule
        \multirow{2}{*}{fs2} & \multirow{2}{*}{Kamei + Coverage} & RF &
        \expTwoRfRocAucValuenum & \expTwoRfPrAucValuenum & \expTwoRfMccValuenum &
        \textbf{\expTwoRfPrecisionValuenum} & \expTwoRfRecallValuenum & \textbf{\expTwoRfFoneValuenum} \\
        & & XGB &
        \expTwoXgbRocAucValuenum & \expTwoXgbPrAucValuenum & \expTwoXgbMccValuenum &
        \textbf{\expTwoXgbPrecisionValuenum} & \expTwoXgbRecallValuenum & \expTwoXgbFoneValuenum \\
        \midrule
        \multirow{2}{*}{fs3} & \multirow{2}{*}{VCCFinder} & RF &
        \expThreeRfRocAucValuenum & \expThreeRfPrAucValuenum & \expThreeRfMccValuenum &
        \expThreeRfPrecisionValuenum & \expThreeRfRecallValuenum & \expThreeRfFoneValuenum \\
        & & XGB &
        \expThreeXgbRocAucValuenum & \expThreeXgbPrAucValuenum & \expThreeXgbMccValuenum &
        \expThreeXgbPrecisionValuenum & \expThreeXgbRecallValuenum & \expThreeXgbFoneValuenum \\
        \midrule
        \multirow{2}{*}{fs4} & \multirow{2}{*}{VCCFinder + Coverage} & RF &
        \textbf{\expFourRfRocAucValuenum} & \expFourRfPrAucValuenum & \textbf{\expFourRfMccValuenum} &
        \textbf{\expFourRfPrecisionValuenum} & \expFourRfRecallValuenum & \textbf{\expFourRfFoneValuenum} \\
        & & XGB &
        \expFourXgbRocAucValuenum & \textbf{\expFourXgbPrAucValuenum} & \expFourXgbMccValuenum &
        \textbf{\expFourXgbPrecisionValuenum} & \textbf{\expFourXgbRecallValuenum} & \textbf{\expFourXgbFoneValuenum} \\
        \midrule
        \multirow{2}{*}{fs5} & \multirow{2}{*}{Coverage} & RF &
        \expFiveRfRocAucValuenum & \expFiveRfPrAucValuenum & \expFiveRfMccValuenum &
        \expFiveRfPrecisionValuenum & \expFiveRfRecallValuenum & \expFiveRfFoneValuenum \\
        & & XGB &
        \expFiveXgbRocAucValuenum & \expFiveXgbPrAucValuenum & \expFiveXgbMccValuenum &
        \expFiveXgbPrecisionValuenum & \expFiveXgbRecallValuenum & \expFiveXgbFoneValuenum \\
        \bottomrule
    \end{tabular}%
    }
\end{table*}
\section{Results}
\label{sec:Result} This section presents the experimental results to answer our three research questions (RQs). First, we evaluate the performance
of our proposed vulnerability prediction model in comparison to baseline approaches (RQ1). Next, we analyze which features are effective for predicting vulnerabilities (RQ2). Finally, we assess the practical value of our approach by evaluating the efficiency of allocating additional fuzzing resources based on vulnerability-risk estimates (RQ3).

\subsection{\rqA}
\noindent\textbf{Approach.}
To answer RQ1, we evaluate the optimal performance achievable by our proposed vulnerability prediction model. Specifically, we follow the cross-project setting detailed in Section~\ref{sec:Method}. For prediction, we use Random Forest and XGBoost models because they have been widely reported to achieve strong performance in JIT bug and vulnerability prediction studies~\cite{DBLP:journals/csur/ZhaoDC23,DBLP:journals/ese/KameiFMYUH16,DBLP:conf/msr/FukushimaKMYU14,DBLP:journals/ese/ZhouHL25,DBLP:journals/sqj/DinterCGT23,DBLP:journals/jss/LomioILPL22}. The feature sets are compared across six configurations: Kamei\etal's features only, VCCFinder’s features only, either of these two feature sets augmented with fuzzing-based dynamic features (denoted as Coverage in Table \ref{tab:full_results}), fuzzing-based dynamic features only, and random prediction as the baseline.

\smallskip\noindent\textbf{Results.}
\autoref{tab:full_results} summarizes the performance of the six configurations (fs0--fs5) with Random Forest and XGBoost classifiers. The best configuration is a Random Forest model trained on the combined VCCFinder and coverage features (fs4), which achieves an average MCC of \expFourRfMccValue, an F1-Score of \expFourRfFoneValue, a Recall of \expFourRfRecallValue, and a PR-AUC of \expFourRfPrAucValue. Its PR-AUC is substantially higher than the random baseline of \expZeroRandomPrAucValue, indicating that the model learned patterns that separate high-risk \ccs{} from lower-risk ones.

Precision is nevertheless limited, so a notable share of warnings would be false positives. This is a well-known challenge in JIT vulnerability prediction: a recent study~\cite{DBLP:journals/corr/abs-2507-10729} reported PR-AUC values of 0.013 for Linux and 0.071 for FFmpeg, and MCC values of 0.030 and 0.122.\footnote{The precision values were not reported in the previous study~\cite{DBLP:journals/corr/abs-2507-10729}.} The low Precision, PR-AUC, MCC, and F1-Score reflect the difficulty of identifying vulnerability-introducing project-days under severe class imbalance, as only \finaldaysratio of the instances in our dataset are positive. Our model thus performs comparably to current state-of-the-art JIT vulnerability prediction models.

To assess the contribution of the fuzzing-based dynamic features, we compare configurations with and without them for each base feature set. Adding these features to VCCFinder (fs3$\rightarrow$fs4) improves the average MCC from \expThreeRfMccValue to \expFourRfMccValue and the PR-AUC from \expThreeRfPrAucValue to \expFourRfPrAucValue; adding them to Kamei\etal's features (fs1$\rightarrow$fs2) improves the average ROC-AUC from \expOneRfRocAucValue to \expTwoRfRocAucValue and the PR-AUC from \expOneRfPrAucValue to \expTwoRfPrAucValue. These improvements are statistically significant. Across the 15 projects, each of the five feature-based configurations (fs1--fs5) significantly outperforms the random baseline on all three primary metrics (Wilcoxon signed-rank test, Benjamini--Hochberg-adjusted $p<0.05$), with large effect sizes for ROC-AUC and MCC (Cliff's $\delta$). Both fs3$\rightarrow$fs4 and fs1$\rightarrow$fs2 also yield significant gains on all three metrics (adjusted $p<0.05$; e.g., $\delta=\deltaFourThreeRocAucValue$ for ROC-AUC on fs3$\rightarrow$fs4). The gains also hold at the project level: PR-AUC improves in 14 of 15 projects for fs3$\rightarrow$fs4 and in all 15 for fs1$\rightarrow$fs2 (median paired differences of 0.032 and 0.028), although the magnitude varies across projects. Fuzzing-based dynamic features thus add predictive information beyond static ones.



\smallskip
\summarybox{\textbf{Answer to RQ1}}{
Our best model (\bestmodel with \bestperfomancefeature features) achieved a mean MCC of \expFourRfMccValue and an F1-Score of \expFourRfFoneValue. It significantly outperformed the random baseline (Wilcoxon signed-rank test, Benjamini--Hochberg-adjusted $p<0.05$; large effect sizes for ROC-AUC and MCC) across all three primary metrics (ROC-AUC, PR-AUC, and MCC).}
\medskip
\subsection{\rqB}
\noindent\textbf{Approach.}
To answer RQ2, we conduct an ablation study to assess the impact of different feature combinations on model performance. Specifically, we compare the results of six experimental configurations (fs0–fs5). 
Additionally, we further investigate the importance of individual features in the best-performing model configuration. For each target project, we extract feature-importance scores (\eg Gini importance for RandomForest) from the trained model and aggregate these scores across all projects. This aggregation allows us to examine the distribution of feature importance and identify the features that are most influential for prediction.

\smallskip\noindent\textbf{Results.}
\autoref{tab:full_results} summarizes the ablation results across the six feature configurations (fs0--fs5).
For MCC, we report only the raw scores, as percentage improvements are not meaningful when the baseline is close to zero or negative.

\smallskip\textit{(a) Comparison of Feature Sets:}
Table~\ref{tab:full_results} shows that the \bestperfomancefeature configuration (\bestperfomanceexp) 
achieves the best overall performance across ROC-AUC, MCC, Precision and F1-Score. These results indicate that combining fuzzing-based dynamic features and traditional features is the most effective approach for vulnerability prediction. A pairwise analysis (paired Wilcoxon test with Benjamini--Hochberg correction) confirms this ranking while clarifying where the gains originate. No other configuration significantly outperforms fs4 on any of the three primary metrics. However, fs4 significantly outperforms Kamei+Coverage (fs2) on ROC-AUC ($\delta=\deltaFourTwoRocAucValue$, adjusted $p<0.05$), while no statistically significant difference was detected for PR-AUC or MCC. This indicates that, once coverage features are included, the choice of static feature set (Kamei vs.\ VCCFinder) has little effect on performance, with \emph{coverage being the dominant signal}, consistent with the feature-importance analysis below. However, adding static features to the coverage-only features (fs5$\rightarrow$fs4) still yields a small but significant improvement for ROC-AUC and MCC (adjusted $p<0.05$), indicating that the two feature families remain complementary. Furthermore, the inferior performance of the configuration that used only fuzzing-based dynamic features (when compared to configurations that used both dynamic and traditional features) highlights the importance of using both static and dynamic features.



\begin{figure}[t]
    \centering
    \includegraphics[width=0.98\columnwidth]{
        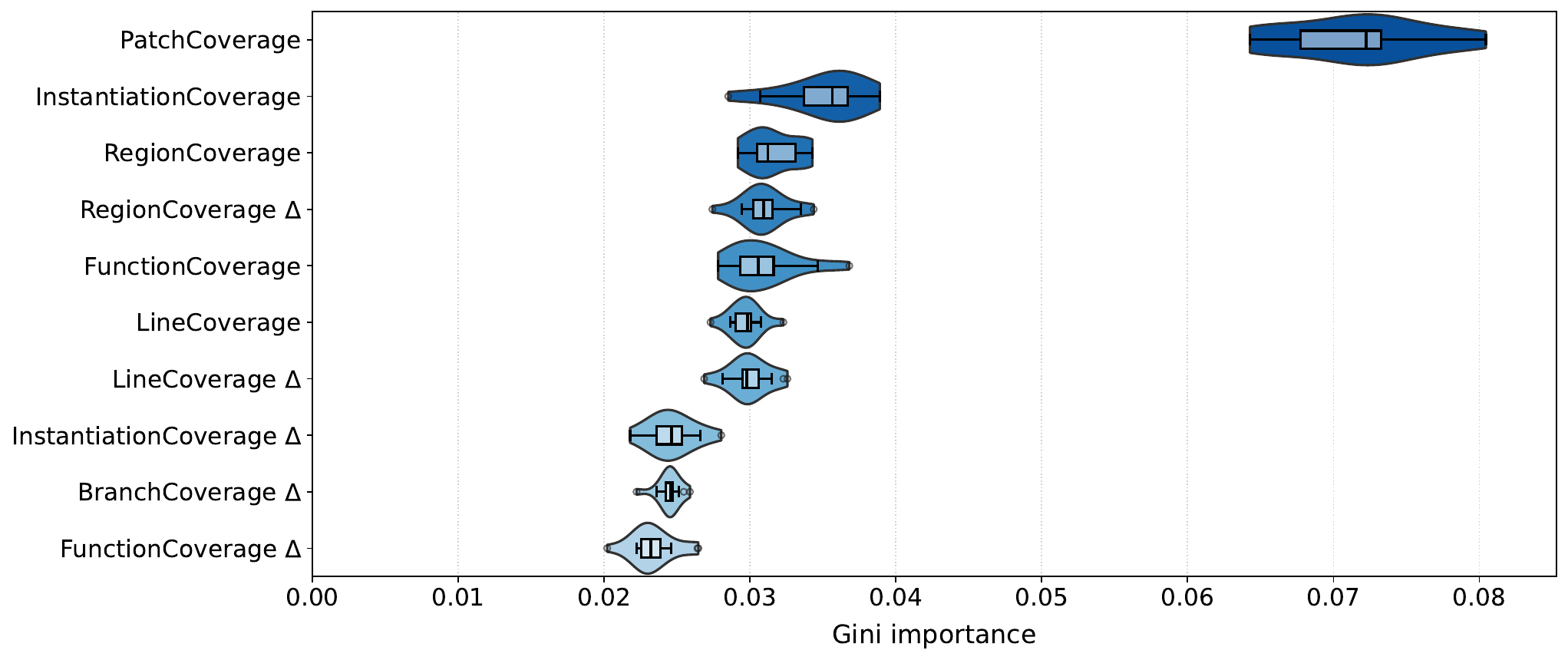
    }
    \caption{Feature importance of the top 10 most important features in the best-performing model (fs4: VCCFinder + Coverage). 
    }
    \label{fig:feature_importance}
\end{figure}

\smallskip\textit{(b) Feature importance of the best model (fs4):}
Figure~\ref{fig:feature_importance} presents the top-10 features for the best-performing model (fs4, \bestmodel). PatchCoverage had the highest feature importance, followed by other coverage metrics. In particular, absolute coverage metrics (\eg PatchCoverage, InstantiationCoverage, RegionCoverage, FunctionCoverage, and LineCoverage) provide a critical baseline for assessing the risk of vulnerability-inducing \ccs.

Several delta-based coverage metrics also appeared among the top-ranked features. Their importance scores were generally lower than PatchCoverage's score, while some delta-based metrics, such as RegionCoverageDelta and LineCoverageDelta, showed importance scores comparable to several absolute coverage metrics. Absolute coverage reflects the extent to which code is exercised by fuzzing, whereas day-to-day coverage fluctuations may indicate changes in testing progress or regression following code changes.

\smallskip
\summarybox{\textbf{Answer to RQ2}}{
Our results show that incorporating fuzzing-derived dynamic features improves vulnerability prediction over static features alone. Coverage-related features, particularly PatchCoverage, are among the most influential features in the best-performing model.
}


\medskip
\subsection{\rqC}
\noindent\textbf{Approach.} 
In RQ3, for each project-day predicted by \APP to contain vulnerability-prone changes in the RQ1 simulation, we apply five strategies to determine to what extent additional fuzzing should be performed.

Because fuzzing time varies across projects due to differences in project size and the number of fuzzing harnesses, we define a \emph{fuzzing session} as one day of fuzzing for a given project. 
For example, if a project is typically fuzzed for 10 minutes per day (the default value in CIFuzz) and a strategy recommends two additional fuzzing sessions, this corresponds to conducting two additional 10-minute fuzzing sessions.

To evaluate the allocation strategies, we must first determine how many fuzzing sessions were required to detect each vulnerability. We use the OSV Database to find the commit when a vulnerability was introduced. We then use the OSS-Fuzz report creation date as the detection date. Since one fuzzing session corresponds to one day, the number of sessions required to detect a vulnerability equals the number of days elapsed between its introduction and detection.

Based on this ground truth, we assess each strategy as follows. If a strategy's suggested number of additional fuzzing sessions equals or exceeds the number of sessions actually required to detect a vulnerability, we consider that the allocation was successful. If the suggested number exceeds the true number required, the excess sessions are treated as wasted. If the suggested number falls short of the true number, all suggested sessions are treated as wasted. Note that if \APP mistakenly predicts a set of changes as likely to be vulnerable, all additional suggested sessions are considered wasted.
Thus, our resource-allocation evaluation explicitly accounts for the cost of both false-positive predictions and over-allocation as wasted fuzzing effort. Since \APP is intended to guide automated fuzzing-resource allocation rather than manual developer inspection, the primary cost of false positives is additional computational effort.

To keep our evaluation rigorous, we applied some constraints. First, in our simulation, each daily prediction is treated independently and does not influence additional fuzzing decisions on subsequent days. We therefore only evaluate vulnerabilities that were introduced on the day being predicted and exclude cases where vulnerabilities from previous days were discovered coincidentally. Second, if no code changes occur on a given day, no additional fuzzing prediction is made.


\begin{table*}[t]
\centering
\caption{Efficiency of Vulnerability Detection Under Simulated Fuzzing-Session Allocation Strategies (RQ3)}
\label{tab:RQ3_summary}
\begin{tabular}{lrrrrrrr}
\toprule
Strategy
  & \makecell[r]{Total Cost \\ (session)}
  & \makecell[r]{Vulnerabilities \\ found}
  & \makecell[r]{Effective Trigger \\ Rate (\%)}
  & \makecell[r]{Cost per \\ vulnerability} \\
\midrule
    S1 Median              & \SoneAdditionalBuildsTotal & \SoneFoundVulns  & \SoneDetectionVulnerabilityRate & \SoneBuildsPerDetection \\
    S2 Random              & \StwoAdditionalBuildsTotal & \textbf{\StwoFoundVulns}  & \textbf{\StwoDetectionVulnerabilityRate} & \StwoBuildsPerDetection \\
    S3 Line-Change Proportional   & \SthreeAdditionalBuildsTotal & \SthreeFoundVulns & \SthreeDetectionVulnerabilityRate & \SthreeBuildsPerDetection \\
    S4 Simple Regression & \SfourAdditionalBuildsTotal & \SfourFoundVulns & \SfourDetectionVulnerabilityRate & \textbf{\SfourBuildsPerDetection} \\
    S5 Multi Regression  & \SfiveAdditionalBuildsTotal & \SfiveFoundVulns & \SfiveDetectionVulnerabilityRate & \SfiveBuildsPerDetection \\
\bottomrule
\end{tabular}
\end{table*}

\smallskip\noindent\textbf{Results.}
Table~\ref{tab:RQ3_summary} summarizes the effectiveness of the five strategies across \numberofvulnerabilities target vulnerabilities.
``Total Cost (Sessions)'' and ``Vulnerabilities Found'' represent the total number of allocated additional fuzzing sessions and the total count of successfully detected vulnerabilities, respectively.
``Cost per Vulnerability'' is calculated by dividing the total cost by the number of vulnerabilities found.
``Effective Trigger Rate'' is defined as the percentage of triggers (\ie executions of additional fuzzing sessions) that successfully identified a vulnerability.
Among the baseline strategies, Median (S1) spends \SoneAdditionalBuildsTotal sessions to find \SoneFoundVulns vulnerabilities at a cost of \SoneBuildsPerDetection sessions per vulnerability (effective trigger rate: \SoneDetectionVulnerabilityRate\%).
Random (S2) increases resource usage to \StwoAdditionalBuildsTotal sessions for \StwoFoundVulns vulnerabilities.
It achieves the highest effective trigger rate (\StwoDetectionVulnerabilityRate\%) because it often allocates more sessions than the other strategies, which increases the chance of reaching the number of sessions required for detection; however, this results in the highest cost per vulnerability at \StwoBuildsPerDetection.
Line-Change Proportional (S3) shows performance comparable to S1, detecting \SthreeFoundVulns vulnerabilities within \SthreeAdditionalBuildsTotal sessions.
Its effective trigger rate is \SthreeDetectionVulnerabilityRate\% and the cost per vulnerability is \SthreeBuildsPerDetection.
The regression-based strategies show different trade-offs.
Simple Regression (S4) has the lowest total cost (\SfourAdditionalBuildsTotal sessions), but also the lowest effective trigger rate (\SfourDetectionVulnerabilityRate\%) and fewer vulnerabilities found (\SfourFoundVulns).
Nevertheless, it achieves the lowest cost per vulnerability (\SfourBuildsPerDetection) among all tested strategies.
Multi Regression (S5) detects \SfiveFoundVulns vulnerabilities with \SfiveAdditionalBuildsTotal sessions, resulting in a cost per vulnerability of \SfiveBuildsPerDetection, comparable to those of S1 and S3.

\begin{figure*}[!t]
    \centering
    \includegraphics[width=0.95\textwidth]{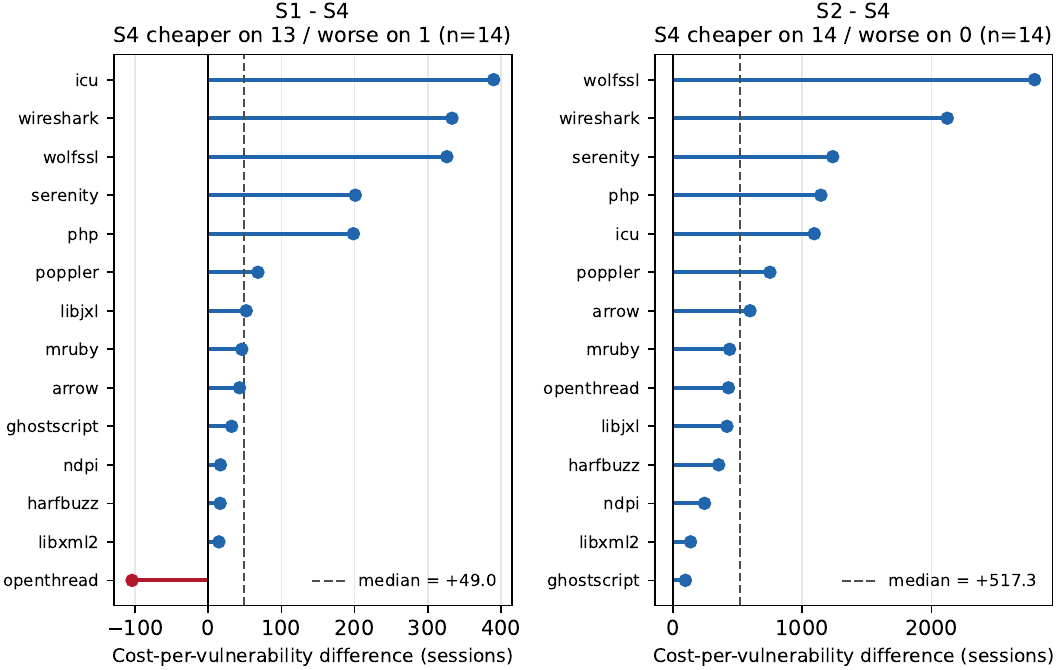}
    \includegraphics[width=0.95\textwidth]{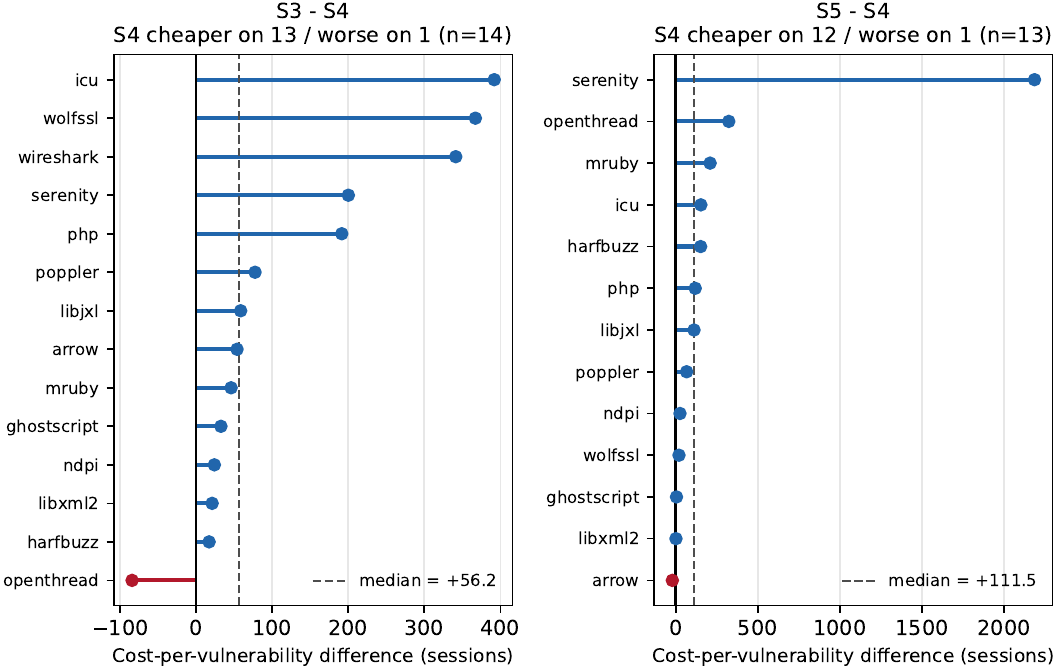}
    \caption{\protect
    Per-project paired differences in cost per vulnerability
    between S4 and each comparison strategy.
    Each point represents the comparison strategy's cost per
    vulnerability minus that of S4 for one project.
    Positive values favor S4, and dashed lines indicate
    median paired differences.
    }
    \label{fig:rq3_paired_cpv_s5}
\end{figure*}

Figure~\ref{fig:rq3_paired_cpv_s5} shows the per-project differences in cost per vulnerability between S4 and the other strategies, with positive values indicating lower costs for S4.
Projects in which either strategy of a pair detected no vulnerability are excluded from that pair, because their cost per vulnerability is undefined.
S4 had a lower cost per vulnerability than S1, S2, S3, and S5 in 13 of 14, 14 of 14, 13 of 14, and 12 of 13 projects, respectively. Thus, the advantage of S4 is consistent across projects. 

To test whether these differences are statistically significant, we compare the strategies on their per-project cost per vulnerability using the paired Wilcoxon signed-rank test with Benjamini--Hochberg correction (Section~\ref{sec:evaluation}).
Simple Regression (S4) is significantly more cost-efficient than each of the three prediction-agnostic heuristics, Median (S1), Random (S2), and Line-Change Proportional (S3) (adjusted $p<0.05$; large effect against S2), and also than Multi Regression (S5) ($\delta=\deltaSFourSFiveValue$, adjusted $p=\pvalSFourSFiveValue$). Multi Regression is significantly more cost-efficient only than Random (S2). The key finding is therefore that \emph{Simple Regression (S4) is the most efficient strategy}, whereas the additional features of Multi Regression do not improve allocation efficiency.
Overall, Simple Regression (S4) is the most cost-efficient strategy, with both the lowest total session cost and the lowest cost per vulnerability.
Random (S2) is the most expensive per vulnerability.



\summarybox{\textbf{Answer to RQ3}}{
    Simple Regression (S4) was the most cost-efficient strategy and was significantly more cost-efficient than the prediction-agnostic heuristics and Multi Regression (S5). S5 did not significantly outperform the Median (S1) and Line-Change Proportional (S3) heuristics. Among all strategies, the Random strategy (S2) was the least cost-efficient, with the highest cost per vulnerability.
}

\medskip
\section{Threats to validity}
In this section, we discuss the potential threats to the validity of our study.
We categorize these threats into three main types: internal validity, construct validity, and external validity. 
\label{sec:Threat}

\smallskip
\textbf{Internal validity:}
A significant threat in cross-project prediction is the heterogeneity of data distributions; metrics such as code size or commit frequency can vary drastically between projects (\eg libraries vs. applications), potentially biasing the model~\cite{DBLP:conf/sigsoft/ZimmermannNGGM09}. 
To address this, we employed tree-based models (Random Forest and XGBoost), which are invariant to monotonic transformations of feature values and thus relatively robust to scale differences across projects~\cite{DBLP:journals/ml/Breiman01,DBLP:conf/kdd/ChenG16}. 


\smallskip
\textbf{Construct validity:}
The accuracy of our ground truth labels relies on the VICs identified via the OSV database~\cite{OSV}. OSV's ``introduced'' field in OSV entries is sometimes recorded as an unknown, potentially introducing noise. To mitigate this threat, we strictly filtered our dataset to include only cases with explicit commit SHAs, ensuring that positive labels are tied to specific code changes.
We formulated the risk prediction at a daily granularity. While this aggregation inevitably poses a risk of signal dilution where benign commits could mask a vulnerability introduced on the same day, our data source, OSS-Fuzz, reports coverage metrics as daily snapshots rather than per-commit. We however believe that this granularity aligns with practical operational workflows in Continuous Fuzzing, where resource allocation decisions (\eg scheduling extended fuzzing campaigns) are typically made on a daily or nightly basis rather than per commit.
Consequently, whether the same findings would hold at commit-level granularity remains unclear, and evaluating this would require per-commit coverage information that is not available in OSS-Fuzz.

Coverage features are obtained from reports generated after the corresponding code changes. In our dataset, 16.1\% of the 968 vulnerabilities were detected on the same day as their introduction, so the corresponding coverage reports may include executions that contributed to the detection itself. This temporal dependency may lead to optimistic estimates of the coverage-aware model's performance.
In RQ3, we also quantified fuzzing effort using the number of daily fuzzing sessions. We acknowledge that this metric simplifies the heterogeneity of actual computational costs (\eg CPU time), which varies by fuzzing target. However, granular resource consumption data (\eg precise CPU hours per fuzzing session) is not publicly available. Consequently, we approximate the amount of additional fuzzing required relative to the usual amount conducted using the number of fuzzing sessions as a proxy for resource allocation.

\smallskip
\textbf{External validity:}
The dataset used in this study is limited to 15 C/C++ projects integrated into OSS-Fuzz.
These projects tend to have a strong focus on security and well-established testing infrastructures. These characteristics and the small number of projects may limit the generalizability of our findings to a broader range of software projects. The predictive utility of the coverage features and the effectiveness of the allocation strategies may also vary across continuous fuzzing environments.
Moreover, the results primarily concern memory corruption vulnerabilities that are characteristic of fuzzing; memory corruption vulnerabilities typically involve unintended modification of memory contents, whereas other types such as injection or improper access control involve exploiting flaws in input handling or authorization logic. It is unclear whether the same feature sets or models would be effective for different classes of vulnerabilities that are prevalent in other languages (\eg injection, improper access control). Therefore, caution is required when applying the findings of this study to other languages or projects.
\section{Conclusion}
\label{sec:Conclusion}
In this paper, we proposed \APP, a \app that uses project-day-level JIT vulnerability-risk estimates to guide the allocation of additional fuzzing resources in continuous fuzzing. Using vulnerability-introducing commits derived from OSV and coverage and change data from OSS-Fuzz, we constructed a daily-level dataset comprising \finalprojects C/C++ projects, \finaltotaldays\xspace project days, and \finalpositivedays vulnerability-introducing days (\finaldaysratio positive rate). On top of this dataset, we empirically evaluated both the predictive performance of \APP and its impact on fuzzing resource allocation.

Our results show that it is possible to identify days on which vulnerabilities are introduced substantially better than random and traditional baselines, although the task remains challenging. 
For RQ1, Random Forest achieves ROC-AUC of up to about \expFourRfRocAucValue and PR-AUC of about \expFourRfPrAucValue, clearly outperforming a random baseline (ROC-AUC = \expZeroRandomRocAucValue, PR-AUC = \expZeroRandomPrAucValue) despite the extreme imbalance in the data.
For RQ2, we find that syntax-based static features, combined with  Fuzzing-based Dynamic Features, consistently yield the best performance. In the Random Forest model, the individual feature importance scores further indicate that coverage-related features are the most influential predictors.
For RQ3, we compared five resource allocation strategies using our newly designed simulation framework over 968 vulnerabilities across \finalprojects target projects. 
The Simple-Regression strategy was significantly more cost-efficient than the prediction-agnostic heuristics, requiring \SfourBuildsPerDetection fuzzing sessions per detected vulnerability.

Future work includes several directions. First, we plan to extend \APP to finer-grained units such as individual commits or fuzz targets, once richer per-commit coverage information becomes available, and to explore more expressive representations of code changes (\eg learned embeddings or graph-based models) that may further improve predictive performance under realistic evaluation protocols.
Second, evaluating \APP in an industrial CI/CD setting where developers interact with risk estimates and resource allocation decisions in real time would provide valuable insights into usability, trust, and organizational impact, and could lead to user-facing tools for risk-aware continuous fuzzing.

\begin{acks}
We gratefully acknowledge the financial support of JSPS KAKENHI grants (JP24K02921, JP25K03100, JP26K23818, JP26H02500), as well as ASPIRE grant (JPMJAP2415), JST CREST grant (JPMJCR23M1, JPMJCR26X7), JST FOREST (JPMJFR252W), and the Inamori Research Institute for
Science for supporting Yasutaka Kamei via the InaRIS Fellowship.
\end{acks}

\bibliographystyle{ACM-Reference-Format}

\bibliography{references}

\end{document}